\documentclass[aps,twocolumn,superscriptaddress,showpacs]{revtex4}
\usepackage{graphicx}
\usepackage{hyperref}
\usepackage{bm,color}
\usepackage{dcolumn}
\usepackage{epsfig,amsbsy,bm,amssymb}
\usepackage{tabu}
\usepackage{tabularx}
\renewcommand*{\[}{\begin{equation}}
\renewcommand*{\]}{\end{equation}}

\newcommand{\np}{\newpage}

\newcommand{\eref}[1] {(\ref{#1})}
\newcommand{\Eref}[1] {Eq.~(\ref{#1})}
\newcommand{\Fref}[1] {Fig. \ref{#1}}

\newcommand{\ra}{\rangle}
\newcommand{\la}{\langle}

\newcommand{\be}{\begin{equation}}
\newcommand{\ee}{\end{equation}}
\newcommand{\br}{\begin{eqnarray*}}
\newcommand{\er}{\end{eqnarray*}}
\newcommand{\ba}{\begin{eqnarray}}
\newcommand{\ea}{\end{eqnarray}}
\newcommand{\bp}{\begin{minipage}}
\newcommand{\ep}{\end{minipage}}
\newcommand{\bt}{\begin{tabular}}
\newcommand{\et}{\end{tabular}}

\renewcommand{\r}{{\bm r}}

\newcommand{\w}{\omega}

\begin{document}
\title{Correlation enhancement of high-order harmonics generation in Xe}

\date{\today}

\author{Alexander W. Bray}

\affiliation{Research School of Physics and Engineering, The
Australian National University, Canberra ACT 0200, Australia}
\author{David Freeman}

\affiliation{Research School of Physics and Engineering, The
Australian National University, Canberra ACT 0200, Australia}
 \author{Sebastian Eckart}

 \affiliation{Institut f\"ur Kernphysik, Goethe-Universit\"at,
   Max-von-Laue-Str. 1, 60438 Frankfurt, Germany}




\author{Anatoli S. Kheifets}
\affiliation{Research School of Physics and Engineering, The
Australian National University, Canberra ACT 0200, Australia}

\begin{abstract}
We consider the process of high-order harmonics generation (HHG) in
the xenon atom enhanced by the inter-shell correlation between the
valence $5p$ and inner $4d$ shells. We derive the HHG spectrum from a
numerical solution of the one-electron time-dependent Schr\"odinger
equation multiplied by the enhancement factor taken as the
photoionization cross-sections ratio calculated with and without the
inter-shell correlation. Such a simplified approach is adequate to
describe the experimental HHG spectrum reported by Shiner {\it et al.}
[Nat. Phys. {\bf 7}, 464 (2011)] generated by a single color IR
laser. Similarly, we find good agreement when  applied to the two-color
driven HHG spectra reported by Faccial\'a {\it et al.}
[Phys. Rev. Lett. {\bf 117}, 093902 (2016)].
%

\end{abstract}

\maketitle

\section{Introduction}

The inter-shell correlation in atomic photoionization is a well
documented phenomenon \cite{A90}. Instead of directly removing an
electron from a designated atomic shell, the photon can be absorbed by
an adjacent, inner or outer, shell. The photoelectron then either
scatters inelastically on the ionized core, or recombines with 
its parent ion.
In the latter, the released energy is 
transferred to remove an electron from said atomic shell. 
Such an inter-shell correlation can modify very
considerably the photoionization cross-section (PICS) in a given
atomic shell. In the Xe atom, the outer $5p$ PICS is significantly
enhanced by the correlation with the inner $4d$ shell near the
so-called giant resonance \cite{Connerade1986}.  This resonance is
formed when the outgoing photoelectron is trapped in a combined field
of the Coulomb potential of the residual ion and the centrifugal
potential. This process leads to enhancement of the $5p$ PICS 
which can exceed an order of magnitude. Not surprisingly, a
similar correlation enhances the HHG spectrum of Xe in the same photon
energy range
\cite{Shiner2011,PhysRevLett.117.093902,0953-4075-51-13-134002}. Indeed,
the photorecombination, the last stage of the three-step picture of the
HHG process \cite{Co94}, is the inverse to photoionization and their
respective cross-sections are related through the principle of
detailed balance.
This principle was first applied in the modelling of HHG by 
\citet{2008PRLLin}
and forms the basis of the Quantitative Rescattering (QRS) 
\cite{2018JPhysBLin} approach.
In which, the spectra is determined by the returning electron wavepacket 
multiplied by the corresponding dipole transition amplitude.
For a comprehensive application of the QRS method to single-color HHG 
from Xe accounting for propagation effects and further experimental 
parameters the readers are referred to \cite{2011PRALin,2012JPhysBLe}.
Under this same factorization, analytic expressions 
for the high-energy end of the HHG plateau were derived by 
\citet{2009PRLStarace,2011PRAStarace,2012PRAStarace}.

Our theoretical treatment of the correlation enhanced HHG
spectrum of Xe is similarly based on this principle. 
First, the HHG spectrum of
Kr was recorded in the spectral range free from resonant enhancement
\cite{Shiner2011,0953-4075-45-7-074010}. Then this spectrum was
normalized to the experimental PICS of the $4p$ shell in the
polarization direction $\sigma_z(4p)=\sigma_{4p}(1+\beta_{4p})$,
$\sigma$ being the total PICS and $\beta$ the angular anisotropy
parameter. The proportionality factor was found to be virtually flat as
a function of the photon energy. Then this factor was multiplied by the
experimental PICS $\sigma_z(5p)$ and compared with the recorded HHG
spectrum of Xe. The comparison proved very convincing. 
Subsequent, more refined, theoretical treatment
within the strong-field approximation \cite{Patchkovskii2012} and the
time-dependent $R$-matrix theory 
\cite{PhysRevA.90.043418,doi:10.1080/09500340.2011.559315} confirmed
this analysis.

In the more recent measurements
\cite{PhysRevLett.117.093902,0953-4075-51-13-134002}, the correlation
enhanced HHG spectrum of Xe was recorded using the two-color $\w/2\w$
IR laser field. This spectrum was systematically studied as a function
of the relative $\w/2\w$ phase which allowed for a tunable enhancement
of a narrow spectral range to be overlapped with the giant resonance
of Xe. Comparison was made with the solution of the time-dependent
Schr\"odinger equation (TDSE) expanded on the specially designed
configuration-interaction basis (the so-called time-dependent
configuration-interaction singles - TDCIS)
\cite{PhysRevLett.111.233005}.  As expected, the HHG spectral range of
the giant resonance was greatly amplified, both by the phase tuning
and the inter-shell correlation. Agreement with the TDCIS calculation
was achieved only when the intershell correlation between the $5p$,
$5s$ and $4d$ shells was taken into account. However, there was a
noticeable disagreement around 63~eV photon energy discussed both in
\cite{PhysRevLett.117.093902} and \cite{0953-4075-51-13-134002} and
attributed to interference between the various photoelectron
trajectories recombining with the parent ion from the opposite
sides. It was concluded that this interference was blurred  in the
experiment  due to a likely spatial and temporal averaging effects.

In the present work, we consider both the single-color and two-color
driven HHG spectra of Xe within the same theoretical model.  We derive
the HHG spectrum from a numerical solution of the one-electron TDSE
multiplied by the enhancement factor. This factor is taken as the
PICS ratio calculated with and without the  $5p/4d$ inter-shell
correlation. We find this approach to work for both the single-color driven
HHG as was previously documented \cite{Shiner2011}, and the two-color field 
with the additional phase dependancy introduced.

\section{Methods}

\subsection{One-electron TDSE and HHG spectrum}

As previously \cite{PhysRevA.97.063404,PhysRevA.98.043427}, we solve
the one-electron TDSE for a target atom
\begin{equation}
\label{TDSE}
i {\partial \Psi(\r,t) / \partial t}=
\left[\hat H_{\rm atom} + \hat H_{\rm int}(t)\right]
\Psi(\r,t) \ ,
\end{equation}
where the radial part of the atomic Hamiltonian
\be
\label{Hat}
\hat H_{\rm atom}(r) = 
-\frac12{d^2\over dr^2} +{l(l+1)\over 2r^2} + V(r)
\ee
contains an effective one-electron potential $V(r)$ 
(Kr \cite{PhysRevA.91.023415}, Xe \cite{PhysRevA.98.043427}). 

The Hamiltonian $\hat H_{\rm
  int}(t)$ describes  interaction with the external field and is
written in the velocity gauge
\be
\label{gauge}
\hat H_{\rm int}(t) =
 {\bm A}(t)\cdot \hat{\bm p} \ \ , \ \ 
{\bm E(t)}=-\partial \bm{A}(t)/\partial t\ .
\ee
This external field is given by%
\be
\label{Two-color}
A_z(t,\alpha,\phi) = -A_{z0}f(t)
[\sin\w t+\alpha\sin(2\w t+\phi)]\ .
\ee
The $f(t)$ is a Gaussian envelope of the form given in \cite{patchkovskii2016simple}.
In the one-color case ($\alpha=0$) we set these parameters to correspond to a 
1800 nm pulse of peak intensity $1.8\times10^{14}$ W/cm$^2$ and FWHM approximately 8.8 fs
with respect to this intensity.
For the two-color case we take the $\w$ (1550 nm) field to be of FWHM 25 fs with
$7\times10^{13}$ W/cm$^2$ peak intensity, and the electric field strength of the $2\w$ (775 nm)
to be 0.4 of that of the primary.

Once the time-dependent wave function $\Psi(r, t)$ is obtained, the
induced dipole moment in the length and acceleration forms can be
calculated \cite{burnett1992calculation}:
\ba
d_L(t) &=& \la\Psi(t)|z|\Psi(t)\ra\\
d_A(t) &=& \la\Psi(t)\left| - dV(r)/dz + E(t)\right|\Psi(t)\ra
\label{Dipole}
\ea
where $z$ is the displacement along the laser polarisaiton axis.
The HHG power spectra can be obtained by the
Fourier transformation of time-dependent dipole
moment, it is expressed as
\ba 
P_L(\w) &\propto&\omega^4 \left|\int\limits_{-T/2}^{T/2}
d_L(t)e^{-i\w t}dt\right|^2 
\\
P_A(\w) &\propto&\phantom{\omega^4}\left|\int\limits_{-T/2}^{T/2}
d_A(t)e^{-i\w t}dt\right|^2 
\label{Fourier}
\ea
for a pulse of length $T$.
Typically the dipole acceleration form, when applicable, 
is preferred as it is not influenced by flux far from
the origin, and hence is more rapidly convergent.
This is the case for our calculations and accordingly, what we use.

\begin{figure}[htbp] \centering
  \includegraphics[width=8.0cm]{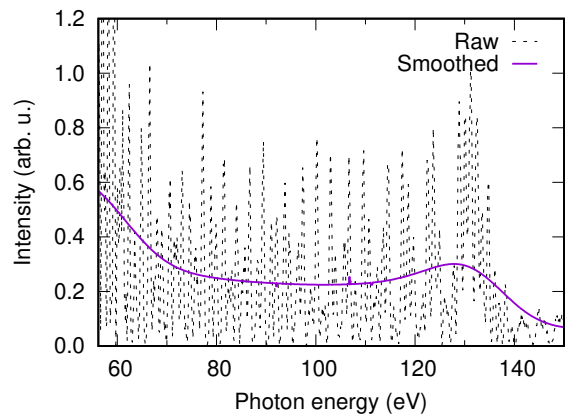}

	\caption{(Color online) Comparison between the raw theoretical 
	HHG spectum \eref{Fourier} (dashed black line) and that 
	smoothed via Gaussian convolution \eref{Conv} (blue solid line). 
	The spectra in particular are for the single color case on Kr.}
\label{FigNew}
\end{figure}

It should also be noted that theoretical spectra as derived above 
exhibit sharp oscillations over several orders of magnitude at an 
energy resolution unattainable by experiment.
Consequently, for comparison, filtering techniques are applied to 
the theoretical spectrum designed to mimic experimental detection.
For this purpose we choose Gaussian convolution of the form
\ba
P^S(\w)&=\frac{\int_0^\infty P(\w')e^{-\sigma(\w-\w')^2} d\w'}
{\int_0^\infty e^{-\sigma(\w-\w')^2} d\w'}\ ,
\label{Conv}
\ea
with $\sigma=0.01$. In practice, the upper integration bound is the 
highest photon energy in the theoretical spectrum.
The effect of such convolution is demonstrated for our single color 
Kr calculation in \Fref{FigNew}.

It should also be noted that comparison with experiment 
is further complicated by macroscopic propagation
effects through the medium whereas theory typically provides only the 
single atom response.
Accordingly however, experiments are conducted with thin 
targets to minimize such macroscopic effects \cite{Shiner2011}.

\subsection{Inter-shell correlation}

\begin{figure}[htbp] \centering
  \includegraphics[width=8.0cm] {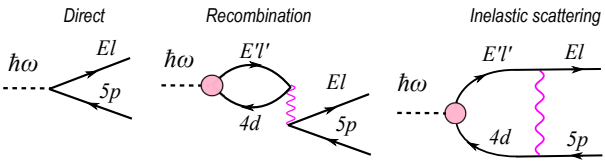} 
  \caption{(Color online) Graphical representation of the direct
    (left) and correlation (center and right) photoionization
    processes included in the RPAE. Straight lines with
    arrows to the right/left represent a photoelectron/hole. The dashed
    line exhibits a photon and the wavy line displays the Coulomb
    interaction. The dashed circle indicates summation of direct and
    exchange correlation processes to infinite order. The arrow
    direction corresponds to photoionization and is the reverse
    in the case of HHG.}
\label{Fig1}
\end{figure}

Random phase approximation with exchange (RPAE) has long been used to
treat inter-shell correlations in noble gas atoms \cite{A90}. In this
approximation, the two inter-shell correlation processes are
considered: recombination and inelastic electron scattering. These 
processes contain the direct and exchange matrix elements of the
Coulomb interaction which is highlighted in the acronym RPAE. The
photoionization amplitude calculated in the RPAE contains the three
terms shown graphically in \Fref{Fig1}. 

In the $5p/4d$ intershell correlation of Xe considered here, the
direct term corresponds to the removal of the electron from the valence
shell and creation of the $5p$ hole. In the recombination process, the
photon is absorbed by the inner shell and the $4d$ hole is created
initially. Then this hole recombines with the photoelectron and the
Coulomb interaction ejects an electron from the outer $5p$ shell. In
the inelastic scattering process, the photoelectron scatters on the
$4d$ hole and promotes it to the $5p$ state. Thus the final state in
all three processes is identical and the corresponding photoionization
amplitudes should be added coherently. The direct photoionization
amplitude in \Fref{Fig1} contains the matrix element of the 
dipole electromagnetic operator acting on the Hartree-Fock
bound and continuous states. The correlation amplitudes contain an
infinite sum of the matrix elements driven by the dipole
electromagnetic operator and containing the Coulomb interaction to
infinite order. 

Relativistic extension of random phase approximation (RRPA)
\cite{PhysRevA.20.964} has been developed to treat heavier atoms. It
accounts for the same direct and exchange correlation processes
exhibited in \Fref{Fig1} but the acronym RRPA does not highlight the
exchange for brevity. The relativistic basis of the one-electron
states contains the Dirac-Fock orbitals, bound and continuous.

We define the $5p/4d$ correlation enhancement factor as the ratio of
the polarization direction PICS calculated with and without
inter-shell correlation. Both the PICS and $\beta$ parameters are
calculated in the RPAE and RRPA by summation of all the three
amplitudes shown in \Fref{Fig1}.  While the correlation-free PICS and
$\beta$ are obtained from the direct amplitude only.

For completeness, we also consider the correlation enhanced
photoionization of Xe calculated within the relativistic time-dependent
density functional theory (RTDDFT) \cite{0953-4075-35-5-313}.
The correlation free calculation in this work is performed using the
relativistic Kohn-Sham (RKS) orbitals. Both the PICS and the $\beta$
parameters are presented in the RTDDFT and RKS and hence the
polarization direction PICS can be evaluated both with and without
inter-shell correlation.

\section{Results and discussion}

\subsection{Enhancement factor}

\begin{figure}[htbp] \centering
  \includegraphics[width=8.0cm]{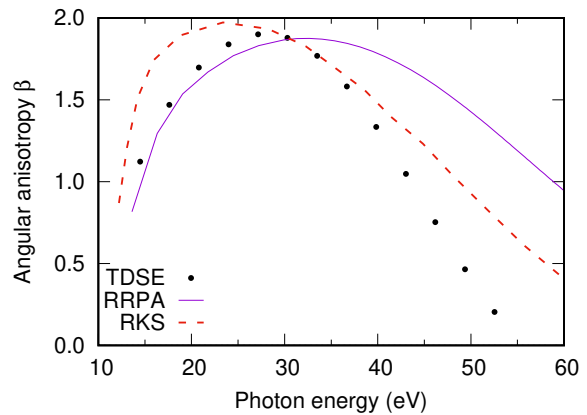}

  \caption{(Color online) The angular anisotropy parameters ($\beta$) resulting from
  the harmonic peaks of our TDSE (black points), 
  and RRPA calculations (blue solid line), 
  and from RKS \cite{0953-4075-35-5-313} (red dashed line). }
\label{Fig2}
\end{figure}

The correlation enhancement factor is calculated as the PICS ratio
\be
\rho(\w) = 
\sigma^{5p/4d}_z(\w)
/
\sigma^{5p}_z(\w)
\ ,
\label{enhancefactor}
\ee
where $\sigma^{5p/4d}_z$ refers to the polarization direction PICS
calculated with the account for the $5p/4d$ inter-shell correlation
whereas $\sigma^{5p}_z(\w)$ is calculated without such a
correlation. The non-correlated RRPA, RPAE and RTDDFT calculations
employ different one-electron basis sets. These bases are not
necessarily commensurate with the one-electron TDSE. Even though we
selected the one-electron potential in \Eref{TDSE} carefully, an
additional test is needed to make sure we can apply various
correlation enhancement factors to the TDSE generated HHG spectrum. In
this test, we compare the angular anisotropy $\beta$ parameters
derived from non-correlated RRPA and RKS calculations. 
We further compare with the parameters resulting from our choice of 
Xe pseudopotential (labelled TDSE)
obtained from the angular dependence of the photoelectron spectrum 
as described in
\cite{PhysRevA.97.063404}. Results of all these calculations are shown
in \Fref{Fig2}. Even though numerically the $\beta$
parameters differ, the general trend of their energy dependence is
very similar in all the calculations. 

\begin{figure}[htbp] \centering
  \includegraphics[width=8.0cm]{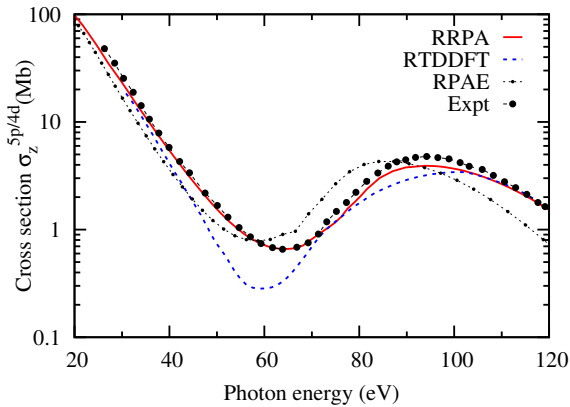}

  \caption{(Color online) The polarization direction photoionization cross-section (PICS)
    $\sigma_z(5p)=\sigma_{5p}(1+\beta_{5p})$. The (red) solid line -
    our RRPA calculation, the (blue) dashed line - RTDDFT \cite{0953-4075-35-5-313},
    the dot-dashed line - our RPAE calculation,
    the filled circles - experimental data compiled in
    \cite{Shiner2011} and renormalized to RRPA.}
\label{Fig3}
\end{figure}

In \Fref{Fig3} we present the polarization direction PICS enhanced by
the correlation  $\sigma^{5p/4d}_z$ returned by different
calculations, RRPA, RPAE and RTDDFT, and compare them with the
experimental values collated in \cite{Shiner2011}. It appears that the
RRPA calculation is closest to the experiment.

\begin{figure}[htbp] \centering

  \includegraphics[width=8.0cm]{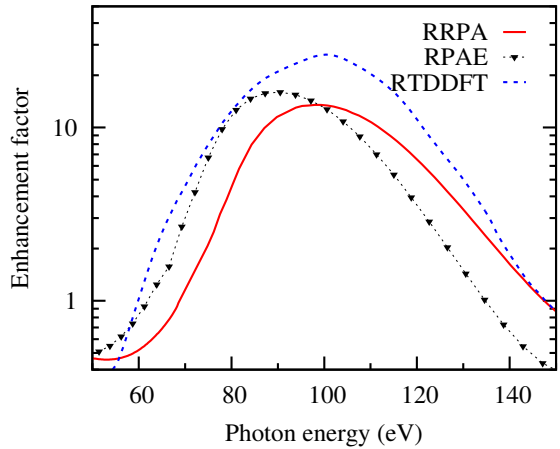}

  \caption{(Color online) The correlation enhancement factor \Eref{enhancefactor}
  from our RRPA calculation (red solid line), 
  our RPAE calculation (black dotted line with triangles), and the
  RTDDFT (blue dashed line) \cite{0953-4075-35-5-313}. }
\label{Fig4}
\end{figure}

Finally, we proceed with the correlation enhancement factor
calculation. This factor is displayed in \Fref{Fig4}. 

\subsection{Single color}

\begin{figure}[htbp] \centering

  \includegraphics[width=8.0cm]{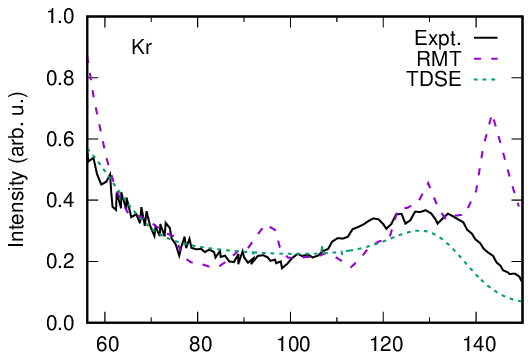}\\
  \includegraphics[width=8.0cm]{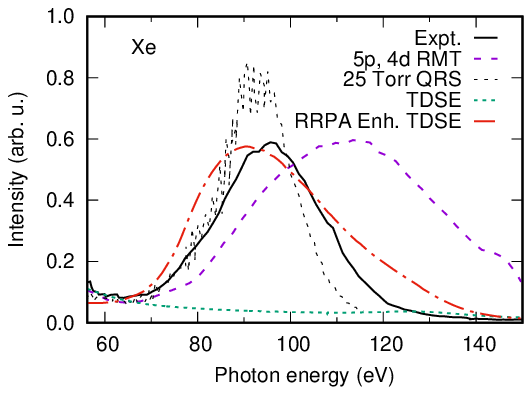}

  \caption{(Color online) Top: The HHG spectrum
    of Kr from experiment \cite{Shiner2011} (black solid line),
    $R$-matrix calculation \cite{PhysRevA.90.043418} (blue sparse dashed line),
    and our present TDSE theory (green dense dashed line).
    Bottom: As above for Xe with the addition of our RRPA enhanced TDSE theory 
	(red dash-dotted line) and QRS calculation \cite{2012JPhysBLe} (black thin dashed line). }
\label{Fig5}
\end{figure}

As the first test of our computational procedure we calculate the HHG
spectrum of Kr, as is unaffected by the correlation
enhancement. We make a comparison with the experiment
\cite{Shiner2011,0953-4075-45-7-074010} and the $R$-matrix calculation
\cite{PhysRevA.90.043418}.  
The calculated spectrum derived from
Eqs.~\eref{TDSE}, \eref{Dipole} and \eref{Fourier} and smoothed by a
Gaussian convolution \eref{Conv} is shown by the green dashed line in the 
top frame of \Fref{Fig5}. 
This compares favourably both with the experiment
\cite{Shiner2011} (green line) and against an analogous $R$-matrix calculation 
\cite{PhysRevA.90.043418} (blue dashed line). 
This is in particularly true for the location of the spectral cut-off albeit the 
relative magnitude of the experiment is marginally higher.
All three datasets have approximately equal predictions for the Cooper minimum.

To test our enhancement procedure, we calculate single color HHG spectrum from Xe. 
Our results are given in the bottom frame of \Fref{Fig5} as enhanced (red dot-dashed line) 
and otherwise (green dashed line) and compared again with experiment \cite{Shiner2011} 
(black solid line) and $R$-matrix calculation \cite{PhysRevA.90.043418} (blue dashed line). 
Additionally we include the QRS 25 Torr dataset from Fig.\ 3c of \cite{2012JPhysBLe} given 
for peak intensity $2\times 10^{14}$ W/cm$^2$.
This latter approach includes macroscopic effects introducing dependencies on the focusing position
and gas pressure.
Of the QRS calculations presented \cite{2011PRALin,2012JPhysBLe} 
it is this which appears to have best agreement with the 
Shiner experiment.
Our reproduction of the giant resonance is an improvement over the RMT 
method. However, both of these approaches exhibit a broader tail at high energies than that seen in 
the experiment. 
Conversely, the QRS method places the resonance well in-line with the experiment but with an altogether
sharper dependency.



\subsection{Two color}

\begin{figure}[htbp] \centering

  \includegraphics[width=8.0cm]{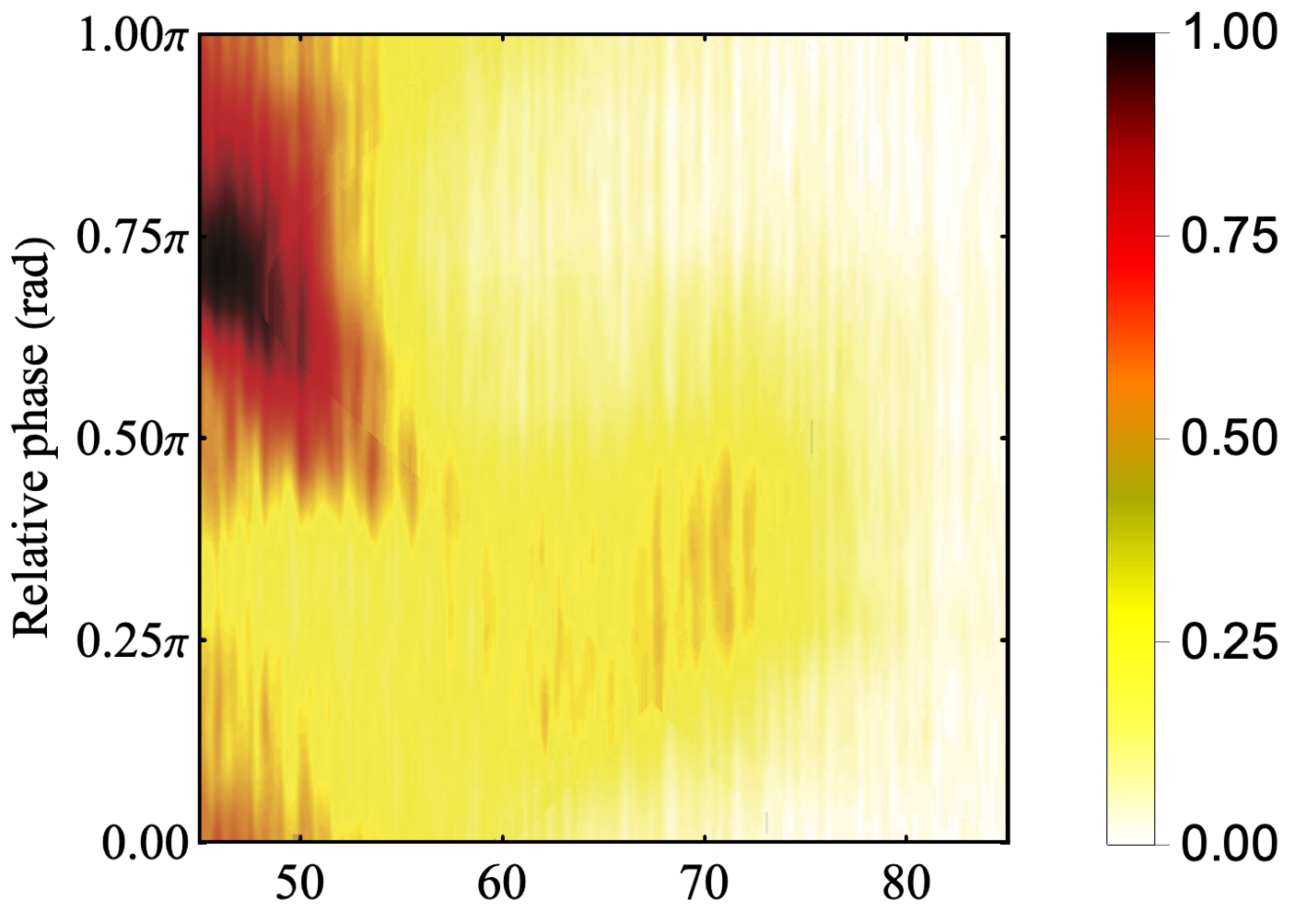}\\
  \hspace{-0.08cm}\includegraphics[width=8.0cm]{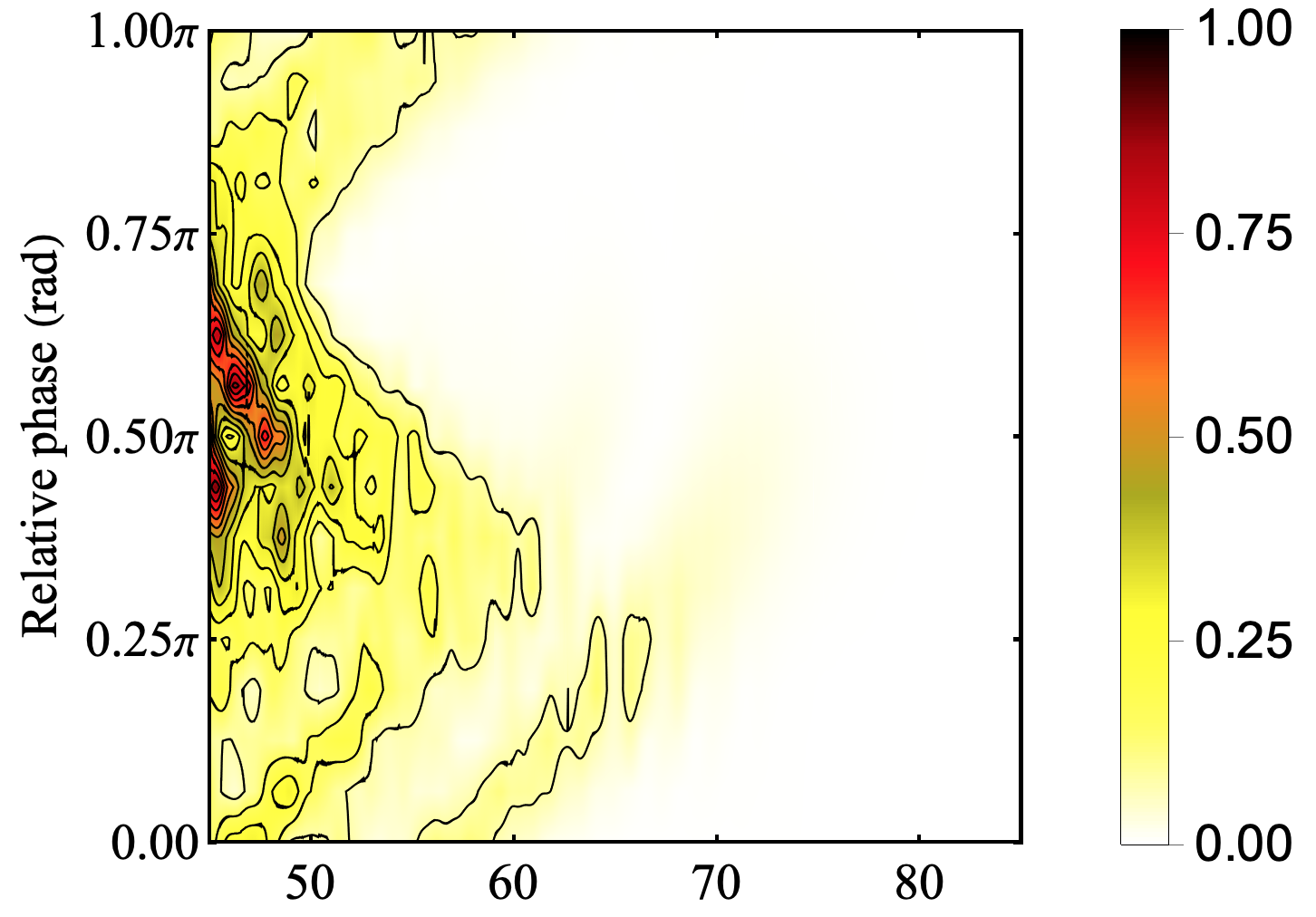}\\
  \includegraphics[width=8.0cm]{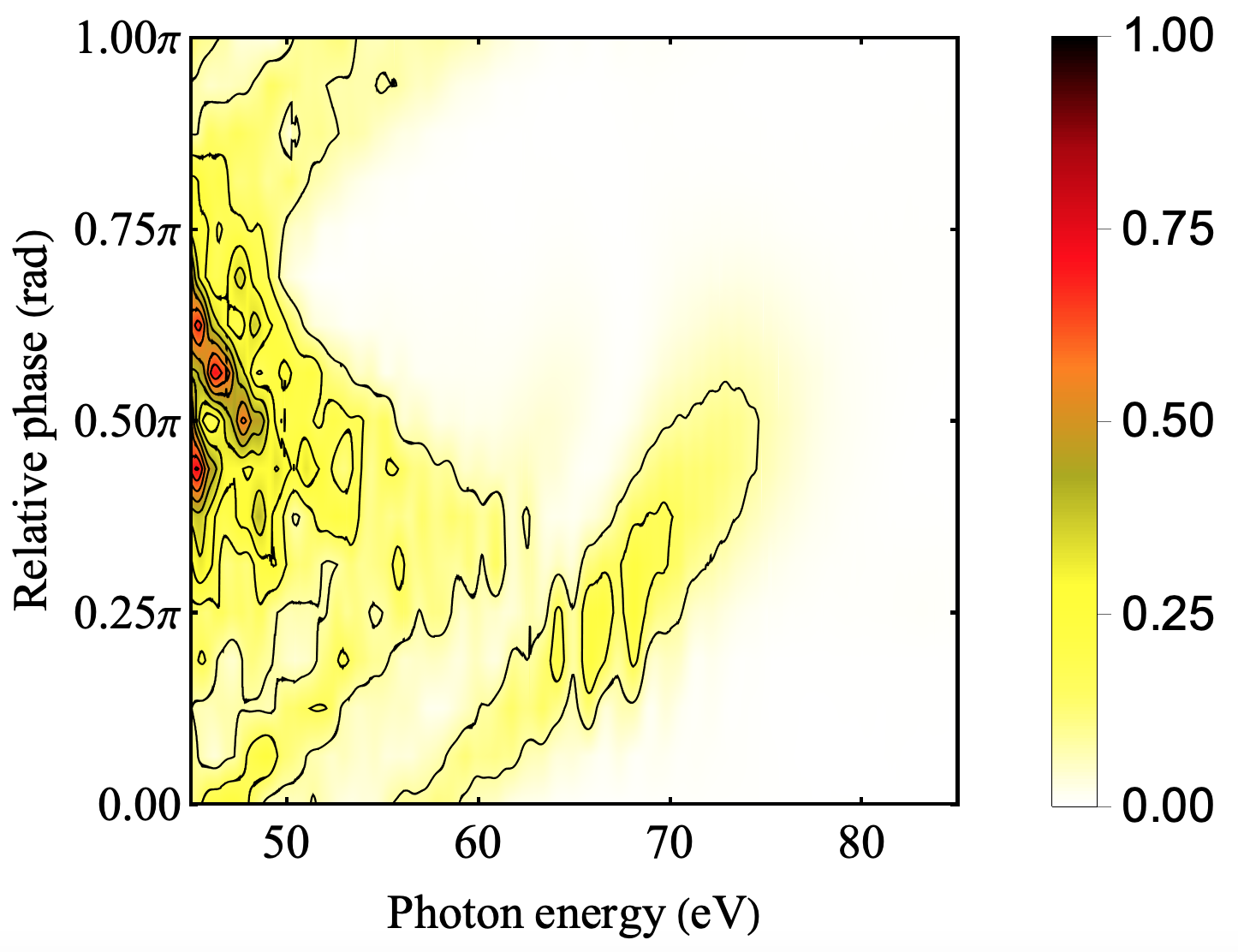}

  \caption{(Color online) The HHG spectrum of Xe recorded as a
    function of the photon energy and the relative phase $\phi$ in \Eref{Two-color}.
 Top: Experimental spectra from \cite{0953-4075-51-13-134002}.
 Middle: Present TDSE calculation.
 Bottom: TDSE spectra enhanced by RRPA. 
 Contours connect regions of equal intensity in intervals of approximately 7\%.}
\label{Fig6}
\end{figure}

We now look to calculate the two-color HHG spectra of Xe presented 
in \cite{PhysRevLett.117.093902}.
The present TDSE calculated  HHG spectra, both raw and enhanced by the RRPA $5p/4d$ 
intershell correlation, are
displayed in the middle and bottom frames of \Fref{Fig6} respectively. 
They are drawn as 2D false color plots in
the energy and relative phase $\phi$ coordinates. The top frame of
\Fref{Fig6} exhibits the experiment from \cite{0953-4075-51-13-134002}. 
We find solid qualitative agreement between our results and that of the experiment, and
as desired, applying the enhancement ratio appropriately 
magnifies the caustic in the cut-off region.
The most notable difference is the differing phase at which the low energy maxima occurs, 
being approximately $\phi=0.75\pi$ in the experiment and our theory predicting closer to 
$0.10\pi$ below this. 
Additionally the experiment exhibits a much stronger relative background across phases than
what we see in our calculation. This being a feature similarly found in their  
comparison with TDCIS theory \cite{PhysRevLett.117.093902}.

\begin{figure}[htbp] \centering

  \includegraphics[width=8.0cm]{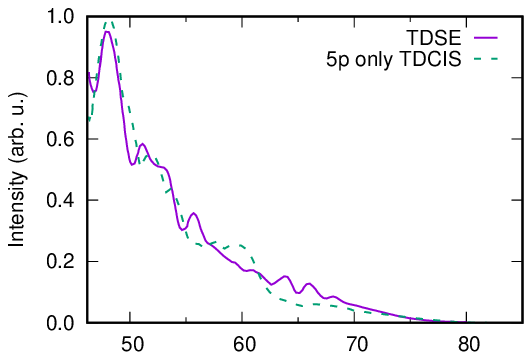}\\
  \includegraphics[width=8.0cm]{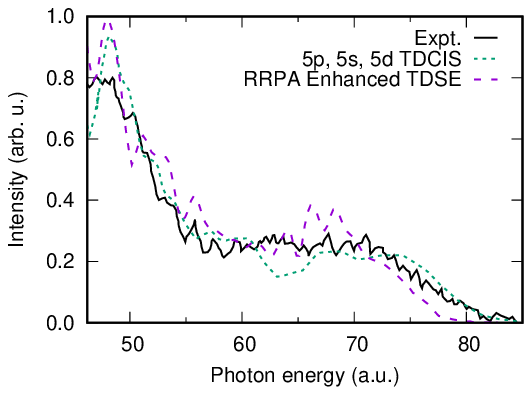}

  \caption{(Color online) 
	The maximum HHG yield of Xe with phase for a given photon energy ($\max_\phi P_A$). 
Top: Our raw  TDSE compared with the 5p only TDCIS theory \cite{PhysRevLett.117.093902}. 
Bottom: Enhanced TDSE result compared with the correlated TDCIS theory 
and experimental measurements 
\cite{PhysRevLett.117.093902}. }
\label{Fig7}
\end{figure}

A more detailed comparison is made in \Fref{Fig7}
where the phase maximum of the spectra shown in \Fref{Fig6} is traced ($\max_\phi P_A$).
On the top panel we display the TDSE calculation compared with the uncorrelated (5p only) 
TDCIS theory. Here we find the observed trend to be essentially equivalent bar minor 
features.
On the bottom panel comparison is made of between our RRPA enhanced spectra and the correlated 
TDCIS (5p, 4d, 5s) as well as the experiment.
Again we find very solid agreement between all three datasets. Most
notably, we did not find a noticeable disagreement between our
calculation and the experiment around 63~eV photon energy which is
visible in the TDCIS calculation.

%
%
%
%
%

\section{Conclusions}

We performed simulations of HHG spectra from Xe in both one- and
two-color fields from the numerical solution of the one-electron TDSE
and account for the effects of intershell correlation as a
ratio of photoionization cross-sections.  In both cases we find strong
agreement between our results and those of experiment and to either be
an improvement, or equivalently accurate, to those of more advanced
theories.  This is attributed to the equivalence of the recombination
process in the production of HHG to the time reversal of
photoionization and accordingly find the correlation effects to behave
similarly, even when the dynamics is complicated significantly by the
introduction of the secondary field. Such an observation suggests
further aspects of two-color HHG production can be efficiently and
effectively studied through the lens of correlated photoionization.

To this end, we will next apply our techniques to other atomic
systems where the HHG process is enhanced by inter-shell correlation
and giant resonances. One such system is atomic manganese where a
giant autoionization resonance due to transition from the $3p$ to a
partially filled $3d$ shell enhances strongly photoionization and
photorecombination from the outer valence shell. This enhancement is
recorded in the HHG spectrum accordingly
\cite{PhysRevLett.121.023201}. 

\section*{Acknowledgements}

The authors are  greatly indebted to Serguei Patchkovskii who
placed his iSURF TDSE code at their disposal. They are also thankful
to Soumyajit Saha for his help with the RRPA calculations and to
XuanYang Lai for many stimulating discussions. 
Resources of the National Computational Infrastructure (NCI Australia)
were employed.

\np~~\np


\begin{thebibliography}{26}
\expandafter\ifx\csname natexlab\endcsname\relax\def\natexlab#1{#1}\fi
\expandafter\ifx\csname bibnamefont\endcsname\relax
  \def\bibnamefont#1{#1}\fi
\expandafter\ifx\csname bibfnamefont\endcsname\relax
  \def\bibfnamefont#1{#1}\fi
\expandafter\ifx\csname citenamefont\endcsname\relax
  \def\citenamefont#1{#1}\fi
\expandafter\ifx\csname url\endcsname\relax
  \def\url#1{\texttt{#1}}\fi
\expandafter\ifx\csname urlprefix\endcsname\relax\def\urlprefix{URL }\fi
\providecommand{\bibinfo}[2]{#2}
\providecommand{\eprint}[2][]{\url{#2}}

\bibitem[{\citenamefont{Amusia}(1990)}]{A90}
\bibinfo{author}{\bibfnamefont{M.~Y.} \bibnamefont{Amusia}},
  \emph{\bibinfo{title}{Atomic photoeffect}} (\bibinfo{publisher}{Plenum
  Press}, \bibinfo{address}{New York}, \bibinfo{year}{1990}).

\bibitem[{\citenamefont{Connerade et~al.}(1986)\citenamefont{Connerade, Esteva,
  and Karnatak}}]{Connerade1986}
\bibinfo{editor}{\bibfnamefont{J.~P.} \bibnamefont{Connerade}},
  \bibinfo{editor}{\bibfnamefont{J.~E.} \bibnamefont{Esteva}},
  \bibnamefont{and} \bibinfo{editor}{\bibfnamefont{R.}~\bibnamefont{Karnatak}},
  eds., \emph{\bibinfo{title}{Giant Resonance in Atoms, Molecules and Solids}}
  (\bibinfo{publisher}{Plenum}, \bibinfo{address}{New York},
  \bibinfo{year}{1986}), no. \bibinfo{number}{151} in \bibinfo{series}{Nato
  Science Series B}.

\bibitem[{\citenamefont{Shiner et~al.}(2011)\citenamefont{Shiner, Schmidt,
  Trallero-Herrero, Worner, Patchkovskii, Corkum, Kieffer, Legare, and
  Villeneuve}}]{Shiner2011}
\bibinfo{author}{\bibfnamefont{A.~D.} \bibnamefont{Shiner}},
  \bibinfo{author}{\bibfnamefont{B.~E.} \bibnamefont{Schmidt}},
  \bibinfo{author}{\bibfnamefont{C.}~\bibnamefont{Trallero-Herrero}},
  \bibinfo{author}{\bibfnamefont{H.~J.} \bibnamefont{Worner}},
  \bibinfo{author}{\bibfnamefont{S.}~\bibnamefont{Patchkovskii}},
  \bibinfo{author}{\bibfnamefont{P.~B.} \bibnamefont{Corkum}},
  \bibinfo{author}{\bibfnamefont{J.~C.} \bibnamefont{Kieffer}},
  \bibinfo{author}{\bibfnamefont{F.}~\bibnamefont{Legare}}, \bibnamefont{and}
  \bibinfo{author}{\bibfnamefont{D.~M.} \bibnamefont{Villeneuve}},
  \emph{\bibinfo{title}{Probing collective multi-electron dynamics in xenon
  with high-harmonic spectroscopy}}, \bibinfo{journal}{Nat Phys}
  \textbf{\bibinfo{volume}{7}}, \bibinfo{pages}{464} (\bibinfo{year}{2011}),
  \bibinfo{note}{10.1038/nphys1940}.

\bibitem[{\citenamefont{Faccial\`a et~al.}(2016)\citenamefont{Faccial\`a,
  Pabst, Bruner, Ciriolo, De~Silvestri, Devetta, Negro, Soifer, Stagira,
  Dudovich et~al.}}]{PhysRevLett.117.093902}
\bibinfo{author}{\bibfnamefont{D.}~\bibnamefont{Faccial\`a}},
  \bibinfo{author}{\bibfnamefont{S.}~\bibnamefont{Pabst}},
  \bibinfo{author}{\bibfnamefont{B.~D.} \bibnamefont{Bruner}},
  \bibinfo{author}{\bibfnamefont{A.~G.} \bibnamefont{Ciriolo}},
  \bibinfo{author}{\bibfnamefont{S.}~\bibnamefont{De~Silvestri}},
  \bibinfo{author}{\bibfnamefont{M.}~\bibnamefont{Devetta}},
  \bibinfo{author}{\bibfnamefont{M.}~\bibnamefont{Negro}},
  \bibinfo{author}{\bibfnamefont{H.}~\bibnamefont{Soifer}},
  \bibinfo{author}{\bibfnamefont{S.}~\bibnamefont{Stagira}},
  \bibinfo{author}{\bibfnamefont{N.}~\bibnamefont{Dudovich}},
  \bibnamefont{et~al.}, \emph{\bibinfo{title}{Probe of multielectron dynamics
  in xenon by caustics in high-order harmonic generation}},
  \bibinfo{journal}{Phys. Rev. Lett.} \textbf{\bibinfo{volume}{117}},
  \bibinfo{pages}{093902} (\bibinfo{year}{2016}).

\bibitem[{\citenamefont{Faccial\`a et~al.}(2018)\citenamefont{Faccial\`a,
  Pabst, Bruner, Ciriolo, Devetta, Negro, Geetha, Pusala, Soifer, Dudovich
  et~al.}}]{0953-4075-51-13-134002}
\bibinfo{author}{\bibfnamefont{D.}~\bibnamefont{Faccial\`a}},
  \bibinfo{author}{\bibfnamefont{S.}~\bibnamefont{Pabst}},
  \bibinfo{author}{\bibfnamefont{B.~D.} \bibnamefont{Bruner}},
  \bibinfo{author}{\bibfnamefont{A.~G.} \bibnamefont{Ciriolo}},
  \bibinfo{author}{\bibfnamefont{M.}~\bibnamefont{Devetta}},
  \bibinfo{author}{\bibfnamefont{M.}~\bibnamefont{Negro}},
  \bibinfo{author}{\bibfnamefont{P.~P.} \bibnamefont{Geetha}},
  \bibinfo{author}{\bibfnamefont{A.}~\bibnamefont{Pusala}},
  \bibinfo{author}{\bibfnamefont{H.}~\bibnamefont{Soifer}},
  \bibinfo{author}{\bibfnamefont{N.}~\bibnamefont{Dudovich}},
  \bibnamefont{et~al.}, \emph{\bibinfo{title}{High-order harmonic generation
  spectroscopy by recolliding electron caustics}}, \bibinfo{journal}{J. Phys.
  B} \textbf{\bibinfo{volume}{51}}(\bibinfo{number}{13}),
  \bibinfo{pages}{134002} (\bibinfo{year}{2018}).

\bibitem[{\citenamefont{Corkum}(1993)}]{Co94}
\bibinfo{author}{\bibfnamefont{P.~B.} \bibnamefont{Corkum}},
  \emph{\bibinfo{title}{Plasma perspective on strong field multiphoton
  ionization}}, \bibinfo{journal}{Phys.~Rev.~Lett.}
  \textbf{\bibinfo{volume}{71}}, \bibinfo{pages}{1994} (\bibinfo{year}{1993}).

\bibitem[{\citenamefont{Morishita et~al.}(2008)\citenamefont{Morishita, Le,
  Chen, and Lin}}]{2008PRLLin}
\bibinfo{author}{\bibfnamefont{T.}~\bibnamefont{Morishita}},
  \bibinfo{author}{\bibfnamefont{A.-T.} \bibnamefont{Le}},
  \bibinfo{author}{\bibfnamefont{Z.}~\bibnamefont{Chen}}, \bibnamefont{and}
  \bibinfo{author}{\bibfnamefont{C.~D.} \bibnamefont{Lin}},
  \emph{\bibinfo{title}{Accurate retrieval of structural information from
  laser-induced photoelectron and high-order harmonic spectra by few-cycle
  laser pulses}}, \bibinfo{journal}{Phys. Rev. Lett.}
  \textbf{\bibinfo{volume}{100}}, \bibinfo{pages}{013903}
  (\bibinfo{year}{2008}).

\bibitem[{\citenamefont{Lin et~al.}(2018)\citenamefont{Lin, Le, Jin, and
  Wei}}]{2018JPhysBLin}
\bibinfo{author}{\bibfnamefont{C.}~\bibnamefont{Lin}},
  \bibinfo{author}{\bibfnamefont{A.-T.} \bibnamefont{Le}},
  \bibinfo{author}{\bibfnamefont{C.}~\bibnamefont{Jin}}, \bibnamefont{and}
  \bibinfo{author}{\bibfnamefont{H.}~\bibnamefont{Wei}},
  \emph{\bibinfo{title}{Elements of the quantitative rescattering theory}},
  \bibinfo{journal}{Journal of Physics B: Atomic, Molecular and Optical
  Physics} \textbf{\bibinfo{volume}{51}}(\bibinfo{number}{10}),
  \bibinfo{pages}{104001} (\bibinfo{year}{2018}).

\bibitem[{\citenamefont{Jin et~al.}(2011)\citenamefont{Jin, Le,
  Trallero-Herrero, and Lin}}]{2011PRALin}
\bibinfo{author}{\bibfnamefont{C.}~\bibnamefont{Jin}},
  \bibinfo{author}{\bibfnamefont{A.-T.} \bibnamefont{Le}},
  \bibinfo{author}{\bibfnamefont{C.~A.} \bibnamefont{Trallero-Herrero}},
  \bibnamefont{and} \bibinfo{author}{\bibfnamefont{C.~D.} \bibnamefont{Lin}},
  \emph{\bibinfo{title}{Generation of isolated attosecond pulses in the far
  field by spatial filtering with an intense few-cycle mid-infrared laser}},
  \bibinfo{journal}{Phys. Rev. A} \textbf{\bibinfo{volume}{84}},
  \bibinfo{pages}{043411} (\bibinfo{year}{2011}).

\bibitem[{\citenamefont{Trallero-Herrero
  et~al.}(2011)\citenamefont{Trallero-Herrero, Jin, Schmidt, Shiner, Kieffer,
  Corkum, Villeneuve, Lin, L{\'e}gar{\'e}, and Le}}]{2012JPhysBLe}
\bibinfo{author}{\bibfnamefont{C.}~\bibnamefont{Trallero-Herrero}},
  \bibinfo{author}{\bibfnamefont{C.}~\bibnamefont{Jin}},
  \bibinfo{author}{\bibfnamefont{B.}~\bibnamefont{Schmidt}},
  \bibinfo{author}{\bibfnamefont{A.}~\bibnamefont{Shiner}},
  \bibinfo{author}{\bibfnamefont{J.}~\bibnamefont{Kieffer}},
  \bibinfo{author}{\bibfnamefont{P.}~\bibnamefont{Corkum}},
  \bibinfo{author}{\bibfnamefont{D.}~\bibnamefont{Villeneuve}},
  \bibinfo{author}{\bibfnamefont{C.}~\bibnamefont{Lin}},
  \bibinfo{author}{\bibfnamefont{F.}~\bibnamefont{L{\'e}gar{\'e}}},
  \bibnamefont{and} \bibinfo{author}{\bibfnamefont{A.-T.} \bibnamefont{Le}},
  \emph{\bibinfo{title}{Generation of broad xuv continuous high harmonic
  spectra and isolated attosecond pulses with intense mid-infrared lasers}},
  \bibinfo{journal}{Journal of Physics B: Atomic, Molecular and Optical
  Physics} \textbf{\bibinfo{volume}{45}}(\bibinfo{number}{1}),
  \bibinfo{pages}{011001} (\bibinfo{year}{2011}).

\bibitem[{\citenamefont{Frolov et~al.}(2009)\citenamefont{Frolov, Manakov,
  Sarantseva, Emelin, Ryabikin, and Starace}}]{2009PRLStarace}
\bibinfo{author}{\bibfnamefont{M.~V.} \bibnamefont{Frolov}},
  \bibinfo{author}{\bibfnamefont{N.~L.} \bibnamefont{Manakov}},
  \bibinfo{author}{\bibfnamefont{T.~S.} \bibnamefont{Sarantseva}},
  \bibinfo{author}{\bibfnamefont{M.~Y.} \bibnamefont{Emelin}},
  \bibinfo{author}{\bibfnamefont{M.~Y.} \bibnamefont{Ryabikin}},
  \bibnamefont{and} \bibinfo{author}{\bibfnamefont{A.~F.}
  \bibnamefont{Starace}}, \emph{\bibinfo{title}{Analytic description of the
  high-energy plateau in harmonic generation by atoms: Can the harmonic power
  increase with increasing laser wavelengths?}}, \bibinfo{journal}{Phys. Rev.
  Lett.} \textbf{\bibinfo{volume}{102}}, \bibinfo{pages}{243901}
  (\bibinfo{year}{2009}).

\bibitem[{\citenamefont{Frolov et~al.}(2011)\citenamefont{Frolov, Manakov,
  Silaev, Vvedenskii, and Starace}}]{2011PRAStarace}
\bibinfo{author}{\bibfnamefont{M.~V.} \bibnamefont{Frolov}},
  \bibinfo{author}{\bibfnamefont{N.~L.} \bibnamefont{Manakov}},
  \bibinfo{author}{\bibfnamefont{A.~A.} \bibnamefont{Silaev}},
  \bibinfo{author}{\bibfnamefont{N.~V.} \bibnamefont{Vvedenskii}},
  \bibnamefont{and} \bibinfo{author}{\bibfnamefont{A.~F.}
  \bibnamefont{Starace}}, \emph{\bibinfo{title}{High-order harmonic generation
  by atoms in a few-cycle laser pulse: Carrier-envelope phase and many-electron
  effects}}, \bibinfo{journal}{Phys. Rev. A} \textbf{\bibinfo{volume}{83}},
  \bibinfo{pages}{021405} (\bibinfo{year}{2011}).

\bibitem[{\citenamefont{Frolov et~al.}(2012)\citenamefont{Frolov, Manakov,
  Sarantseva, and Starace}}]{2012PRAStarace}
\bibinfo{author}{\bibfnamefont{M.~V.} \bibnamefont{Frolov}},
  \bibinfo{author}{\bibfnamefont{N.~L.} \bibnamefont{Manakov}},
  \bibinfo{author}{\bibfnamefont{T.~S.} \bibnamefont{Sarantseva}},
  \bibnamefont{and} \bibinfo{author}{\bibfnamefont{A.~F.}
  \bibnamefont{Starace}}, \emph{\bibinfo{title}{High-order-harmonic-generation
  spectroscopy with an elliptically polarized laser field}},
  \bibinfo{journal}{Phys. Rev. A} \textbf{\bibinfo{volume}{86}},
  \bibinfo{pages}{063406} (\bibinfo{year}{2012}).

\bibitem[{\citenamefont{Shiner et~al.}(2012)\citenamefont{Shiner, Schmidt,
  Trallero-Herrero, Corkum, Kieffer, L.gar., and
  Villeneuve}}]{0953-4075-45-7-074010}
\bibinfo{author}{\bibfnamefont{A.~D.} \bibnamefont{Shiner}},
  \bibinfo{author}{\bibfnamefont{B.~E.} \bibnamefont{Schmidt}},
  \bibinfo{author}{\bibfnamefont{C.}~\bibnamefont{Trallero-Herrero}},
  \bibinfo{author}{\bibfnamefont{P.~B.} \bibnamefont{Corkum}},
  \bibinfo{author}{\bibfnamefont{J.-C.} \bibnamefont{Kieffer}},
  \bibinfo{author}{\bibfnamefont{F.}~\bibnamefont{L.gar.}}, \bibnamefont{and}
  \bibinfo{author}{\bibfnamefont{D.~M.} \bibnamefont{Villeneuve}},
  \emph{\bibinfo{title}{Observation of {Cooper} minimum in krypton using high
  harmonic spectroscopy}}, \bibinfo{journal}{J. Phys. B}
  \textbf{\bibinfo{volume}{45}}(\bibinfo{number}{7}), \bibinfo{pages}{074010}
  (\bibinfo{year}{2012}).

\bibitem[{\citenamefont{Patchkovskii et~al.}(2012)\citenamefont{Patchkovskii,
  Smirnova, and Spanner}}]{Patchkovskii2012}
\bibinfo{author}{\bibfnamefont{S.}~\bibnamefont{Patchkovskii}},
  \bibinfo{author}{\bibfnamefont{O.}~\bibnamefont{Smirnova}}, \bibnamefont{and}
  \bibinfo{author}{\bibfnamefont{M.}~\bibnamefont{Spanner}},
  \emph{\bibinfo{title}{Attosecond control of electron correlations in
  one-photon ionization and recombination}}, \bibinfo{journal}{J. Phys. B}
  \textbf{\bibinfo{volume}{45}}(\bibinfo{number}{13}), \bibinfo{pages}{131002}
  (\bibinfo{year}{2012}).

\bibitem[{\citenamefont{Hassouneh et~al.}(2014)\citenamefont{Hassouneh, Brown,
  and van~der Hart}}]{PhysRevA.90.043418}
\bibinfo{author}{\bibfnamefont{O.}~\bibnamefont{Hassouneh}},
  \bibinfo{author}{\bibfnamefont{A.~C.} \bibnamefont{Brown}}, \bibnamefont{and}
  \bibinfo{author}{\bibfnamefont{H.~W.} \bibnamefont{van~der Hart}},
  \emph{\bibinfo{title}{Harmonic generation by noble-gas atoms in the near-{IR}
  regime using ab initio time-dependent {$R$}-matrix theory}},
  \bibinfo{journal}{Phys. Rev. A} \textbf{\bibinfo{volume}{90}},
  \bibinfo{pages}{043418} (\bibinfo{year}{2014}).

\bibitem[{\citenamefont{Moore et~al.}(2011)\citenamefont{Moore, Lysaght,
  Nikolopoulos, Parker, van~der Hart, and
  Taylor}}]{doi:10.1080/09500340.2011.559315}
\bibinfo{author}{\bibfnamefont{L.}~\bibnamefont{Moore}},
  \bibinfo{author}{\bibfnamefont{M.}~\bibnamefont{Lysaght}},
  \bibinfo{author}{\bibfnamefont{L.}~\bibnamefont{Nikolopoulos}},
  \bibinfo{author}{\bibfnamefont{J.}~\bibnamefont{Parker}},
  \bibinfo{author}{\bibfnamefont{H.}~\bibnamefont{van~der Hart}},
  \bibnamefont{and} \bibinfo{author}{\bibfnamefont{K.}~\bibnamefont{Taylor}},
  \emph{\bibinfo{title}{The {RMT} method for many-electron atomic systems in
  intense short-pulse laser light}}, \bibinfo{journal}{Journal of Modern
  Optics} \textbf{\bibinfo{volume}{58}}(\bibinfo{number}{13}),
  \bibinfo{pages}{1132} (\bibinfo{year}{2011}).

\bibitem[{\citenamefont{Pabst and Santra}(2013)}]{PhysRevLett.111.233005}
\bibinfo{author}{\bibfnamefont{S.}~\bibnamefont{Pabst}} \bibnamefont{and}
  \bibinfo{author}{\bibfnamefont{R.}~\bibnamefont{Santra}},
  \emph{\bibinfo{title}{Strong-field many-body physics and the giant
  enhancement in the high-harmonic spectrum of xenon}}, \bibinfo{journal}{Phys.
  Rev. Lett.} \textbf{\bibinfo{volume}{111}}, \bibinfo{pages}{233005}
  (\bibinfo{year}{2013}).

\bibitem[{\citenamefont{Bray et~al.}(2018{\natexlab{a}})\citenamefont{Bray,
  Naseem, and Kheifets}}]{PhysRevA.97.063404}
\bibinfo{author}{\bibfnamefont{A.~W.} \bibnamefont{Bray}},
  \bibinfo{author}{\bibfnamefont{F.}~\bibnamefont{Naseem}}, \bibnamefont{and}
  \bibinfo{author}{\bibfnamefont{A.~S.} \bibnamefont{Kheifets}},
  \emph{\bibinfo{title}{Simulation of angular-resolved {RABBITT} measurements
  in noble-gas atoms}}, \bibinfo{journal}{Phys. Rev. A}
  \textbf{\bibinfo{volume}{97}}, \bibinfo{pages}{063404}
  (\bibinfo{year}{2018}{\natexlab{a}}).

\bibitem[{\citenamefont{Bray et~al.}(2018{\natexlab{b}})\citenamefont{Bray,
  Naseem, and Kheifets}}]{PhysRevA.98.043427}
\bibinfo{author}{\bibfnamefont{A.~W.} \bibnamefont{Bray}},
  \bibinfo{author}{\bibfnamefont{F.}~\bibnamefont{Naseem}}, \bibnamefont{and}
  \bibinfo{author}{\bibfnamefont{A.~S.} \bibnamefont{Kheifets}},
  \emph{\bibinfo{title}{Photoionization of {Xe} and
  $\mathrm{Xe}@{\mathrm{c}}_{60}$ from the $4d$ shell in {RABBITT} fields}},
  \bibinfo{journal}{Phys. Rev. A} \textbf{\bibinfo{volume}{98}},
  \bibinfo{pages}{043427} (\bibinfo{year}{2018}{\natexlab{b}}).

\bibitem[{\citenamefont{Cloux et~al.}(2015)\citenamefont{Cloux, Fabre, and
  Pons}}]{PhysRevA.91.023415}
\bibinfo{author}{\bibfnamefont{F.}~\bibnamefont{Cloux}},
  \bibinfo{author}{\bibfnamefont{B.}~\bibnamefont{Fabre}}, \bibnamefont{and}
  \bibinfo{author}{\bibfnamefont{B.}~\bibnamefont{Pons}},
  \emph{\bibinfo{title}{Semiclassical description of high-order-harmonic
  spectroscopy of the {Cooper} minimum in krypton}}, \bibinfo{journal}{Phys.
  Rev. A} \textbf{\bibinfo{volume}{91}}, \bibinfo{pages}{023415}
  (\bibinfo{year}{2015}).

\bibitem[{\citenamefont{Patchkovskii and
  Muller}(2016)}]{patchkovskii2016simple}
\bibinfo{author}{\bibfnamefont{S.}~\bibnamefont{Patchkovskii}}
  \bibnamefont{and} \bibinfo{author}{\bibfnamefont{H.}~\bibnamefont{Muller}},
  \emph{\bibinfo{title}{Simple, accurate, and efficient implementation of
  1-electron atomic time-dependent schr{\"o}dinger equation in spherical
  coordinates}}, \bibinfo{journal}{Computer Physics Communications}
  \textbf{\bibinfo{volume}{199}}, \bibinfo{pages}{153} (\bibinfo{year}{2016}).

\bibitem[{\citenamefont{Burnett et~al.}(1992)\citenamefont{Burnett, Reed,
  Cooper, and Knight}}]{burnett1992calculation}
\bibinfo{author}{\bibfnamefont{K.}~\bibnamefont{Burnett}},
  \bibinfo{author}{\bibfnamefont{V.}~\bibnamefont{Reed}},
  \bibinfo{author}{\bibfnamefont{J.}~\bibnamefont{Cooper}}, \bibnamefont{and}
  \bibinfo{author}{\bibfnamefont{P.}~\bibnamefont{Knight}},
  \emph{\bibinfo{title}{Calculation of the background emitted during
  high-harmonic generation}}, \bibinfo{journal}{Physical Review A}
  \textbf{\bibinfo{volume}{45}}(\bibinfo{number}{5}), \bibinfo{pages}{3347}
  (\bibinfo{year}{1992}).

\bibitem[{\citenamefont{Johnson and Lin}(1979)}]{PhysRevA.20.964}
\bibinfo{author}{\bibfnamefont{W.~R.} \bibnamefont{Johnson}} \bibnamefont{and}
  \bibinfo{author}{\bibfnamefont{C.~D.} \bibnamefont{Lin}},
  \emph{\bibinfo{title}{Multichannel relativistic random-phase approximation
  for the photoionization of atoms}}, \bibinfo{journal}{Phys. Rev. A}
  \textbf{\bibinfo{volume}{20}}, \bibinfo{pages}{964} (\bibinfo{year}{1979}).

\bibitem[{\citenamefont{Toffoli et~al.}(2002)\citenamefont{Toffoli, Stener, and
  Decleva}}]{0953-4075-35-5-313}
\bibinfo{author}{\bibfnamefont{D.}~\bibnamefont{Toffoli}},
  \bibinfo{author}{\bibfnamefont{M.}~\bibnamefont{Stener}}, \bibnamefont{and}
  \bibinfo{author}{\bibfnamefont{P.}~\bibnamefont{Decleva}},
  \emph{\bibinfo{title}{Application of the relativistic time-dependent density
  functional theory to the photoionization of xenon}},
  \bibinfo{journal}{J.~Phys.~B}
  \textbf{\bibinfo{volume}{35}}(\bibinfo{number}{5}), \bibinfo{pages}{1275}
  (\bibinfo{year}{2002}).

\bibitem[{\citenamefont{Fareed et~al.}(2018)\citenamefont{Fareed, Strelkov,
  Singh, Thir\'e, Mondal, Schmidt, L\'egar\'e, and
  Ozaki}}]{PhysRevLett.121.023201}
\bibinfo{author}{\bibfnamefont{M.~A.} \bibnamefont{Fareed}},
  \bibinfo{author}{\bibfnamefont{V.~V.} \bibnamefont{Strelkov}},
  \bibinfo{author}{\bibfnamefont{M.}~\bibnamefont{Singh}},
  \bibinfo{author}{\bibfnamefont{N.}~\bibnamefont{Thir\'e}},
  \bibinfo{author}{\bibfnamefont{S.}~\bibnamefont{Mondal}},
  \bibinfo{author}{\bibfnamefont{B.~E.} \bibnamefont{Schmidt}},
  \bibinfo{author}{\bibfnamefont{F.}~\bibnamefont{L\'egar\'e}},
  \bibnamefont{and} \bibinfo{author}{\bibfnamefont{T.}~\bibnamefont{Ozaki}},
  \emph{\bibinfo{title}{Harmonic generation from neutral manganese atoms in the
  vicinity of the giant autoionization resonance}}, \bibinfo{journal}{Phys.
  Rev. Lett.} \textbf{\bibinfo{volume}{121}}, \bibinfo{pages}{023201}
  (\bibinfo{year}{2018}).

\end{thebibliography}

\end{document}